\documentclass[%
twocolumn,
superscriptaddress,
 amsmath,amssymb,
 aps,
prl,
]{revtex4-2}

\usepackage{graphicx}
\usepackage{dcolumn}
\usepackage{bm}

\usepackage{color}

\begin{document}

\preprint{APS/123-QED}

\title{Slow-Wave Hybrid Magnonics}

\author{Jing Xu}
\affiliation{ 
    Center for Nanoscale Materials, Argonne National Laboratory, Lemont, IL 60439, USA
}

\author{Changchun Zhong}
\affiliation{ 
    Pritzker School of Molecular Engineering, University of Chicago, Chicago, IL 60637, USA
}

\author{Shihao Zhuang}
\affiliation{ 
    Department of Materials Science and Engineering, University of Wisconsin - Madison, Madison, WI 53706, USA
}

\author{Chen Qian}
\affiliation{ 
    Department of Physics and Astronomy, 
    University of Pennsylvania, Philadelphia, PA, 19104, USA
}

\author{Yu Jiang}
\affiliation{ 
    Department of Electrical and Computer Engineering, Northeastern University, Boston, MA 02115, USA
}

\author{Amin Pishehvar}
\affiliation{ 
    Department of Electrical and Computer Engineering, Northeastern University, Boston, MA 02115, USA
}

\author{Xu Han}
\affiliation{ 
    Center for Nanoscale Materials, Argonne National Laboratory, Lemont, IL 60439, USA
}

\author{Dafei Jin}
\affiliation{
    Department of Physics and Astronomy, University of Notre Dame, Notre Dame, Indiana 46556, USA
}
\affiliation{ 
    Center for Nanoscale Materials, Argonne National Laboratory, Lemont, IL 60439, USA
}

\author{Josep M. Jornet}
\affiliation{ 
    Department of Electrical and Computer Engineering, Northeastern University, Boston, MA 02115, USA
}

\author{Bo Zhen}
\affiliation{ 
    Department of Physics and Astronomy, 
    University of Pennsylvania, Philadelphia, PA, 19104, USA
}

\author{Jiamian Hu}
\affiliation{ 
    Department of Materials Science and Engineering, University of Wisconsin - Madison, Madison, WI 53706, USA
}

\author{Liang Jiang}
\affiliation{ 
    Pritzker School of Molecular Engineering, University of Chicago, Chicago, IL 60637, USA
}

\author{Xufeng Zhang}
\email{xu.zhang@northeastern.edu}
\affiliation{ 
    Department of Electrical and Computer Engineering, Northeastern University, Boston, MA 02115, USA
}

\date{\today}

\begin{abstract}
Cavity magnonics is an emerging research area focusing on the coupling between magnons and photons. Despite its great potential for coherent information processing, it has been long restricted by the narrow interaction bandwidth. In this work, we theoretically propose and experimentally demonstrate a novel approach to achieve broadband photon-magnon coupling by adopting slow waves on engineered microwave waveguides. To the best of our knowledge, this is the first time that slow wave is combined with hybrid magnonics. Its unique properties promise great potentials for both fundamental research and practical applications, for instance, by deepening our understanding of the light-matter interaction in the slow wave regime and providing high-efficiency spin wave transducers. The device concept can be extended to other systems such as optomagnonics and magnomechanics, opening up new directions for hybrid magnonics.
\end{abstract}


\maketitle



Cavity magnonics is an emerging area exploring the utilization of magnons -- quasiparticles describing collective spin excitations known as spin waves -- for coherent information processing \cite{Rameshti_PhysRep_2022,Harder_SSC_2018,Lachance_APE_2019,Bhoi_SSP_2020,YiLi_JAP_2020,Awschalom_IEEETransQuantEng_2021}. Different from conventional magnonics \cite{Serga2010Jun,Chumak2019May}, cavity magnonics focuses on coherent interaction between magnons and cavity photons instead of the magnonic dynamics itself. Such interaction allows coherent information exchange between magnons and photons, leading to coherent phenomena including strong coupling \cite{Huebl_PRL_2013,XufengZhang_PRL_2014,Tabuchi_PRL_2014,Goryachev_PRAppl_2014,LihuiBai_PRL_2015, JustinHou_PRL_2019,YiLi_PRL_2022}, magnetically induced transparency \cite{XufengZhang_PRL_2014}, and unidirectional invisibility \cite{YipuWang_PRL_2019,XufengZhang_PRAppl_2020}, which enable wide applications in quantum transduction \cite{Hisatomi_PRB_2016,NaZhu_Optica_2020}, dark matter detection \cite{Crescini_PRL_2020,Flower_PhysDarkUniv_2019}, and neuromorphic computing \cite{Millet_APL_2021}.

In cavity magnonics, the magnon-photon coupling can be significantly enhanced via photon recycling at resonances. Through proper cavity design, the magnon-photon coupling strength can surpass system dissipation levels, enabling an efficient and robust interaction between the two modes. For instance, at microwave frequencies, metallic three-dimensional cavities or coplanar waveguide (CPW) resonators are often used to host magnonic resonators, allowing the observation of strong \cite{Huebl_PRL_2013,XufengZhang_PRL_2014,Tabuchi_PRL_2014,Goryachev_PRAppl_2014,LihuiBai_PRL_2015,JustinHou_PRL_2019,YiLi_PRL_2022} and even ultrastrong \cite{XufengZhang_PRL_2014,Bourhill_PRB_2016,Golovchanskiy_SciAdv_2021} photon-magnon coupling. When coupling magnons with optical photons or mechanical phonons where the interaction is intrinsically weak, complicated triple-resonance conditions can be applied to achieve further interaction enhancement \cite{XufengZhang_SciAdv_2016, XufengZhang_PRL_2016,Osada_PRL_2016,Haigh_PRL_2016}.

Despite the enhanced coupling, cavity magnonics suffers from its finite bandwidth. Although the magnon frequency is widely tunable, the coherent magnon-photon coupling can only be observed near the fixed cavity frequency. Within the limited exploration of broadband magnon-photon coupling, only cascaded discrete features are obtained \cite{Bhoi2021Feb}, which remains inherently narrow-band. Such small bandwidths complicate device designs and limit applications of cavity magnonics. To break this restriction, non-resonant structures supporting broadband traveling photons are desired. However, the interaction of traveling photons with magnons is usually weak. Particularly, it is extremely challenging to detect magnon signals on integrated devices where micro/nano-magnonic resonators are coupled with microstrips or CPWs, which usually requires sophisticated technologies such as Brillouin light scattering \cite{Sebastian_FrontPhys_2015} that are expensive and incompatible with large-scale device integration. 

To address this challenge, we propose a novel concept of slow-wave hybrid magnonics. It originates from previous demonstrations in optical domain where slow lights -- traveling optical photons with reduced group velocities -- are used to enhance light-matter interactions and enable new functionalities without compromising the bandwidth \cite{Jensen_OptCommun_2008Nov,Thevenaz_SPIE_2012,Zang_PRAppl_2016Feb,Calajo_2017_PRA}. By introducing spoof surface plasmon polariton (SSPP) structures \cite{Pendry_Science_2004,Gao_AdvMater_2018,Tang_AOM_2019,Garcia-Vidal_RMP_2022}, slow waves can also be obtained for microwave photons with largely reduced group velocities. SSPP has experienced rapid development in the past decades and found wide applications in developing compact microwave/THz devices. Here, for the first time, we combine the two promising fields -- spoof plasmonics \cite{Gao_AdvMater_2018} and cavity magnonics -- and show that it can lead to broadband hybrid magnonic interactions while maintaining large coupling strengths, as well as enable complicated system dynamics that is of critical importance for complex magnonic systems such as magnonic crystals \cite{Chumak_JPD_2017May,De_JMMM_2019Oct,De_PRB_2021Feb,Pal_AQT_2023Jun,Adhikari_AANM_2023May} and magnonic networks \cite{Khivintsev2022Mar}. Furthermore, we demonstrate slow-wave strong coupling  within our system, a phenomenon that has not been experimentally observed in magnonics or other systems.

\begin{figure}[tb]
\includegraphics[width=0.99\linewidth]{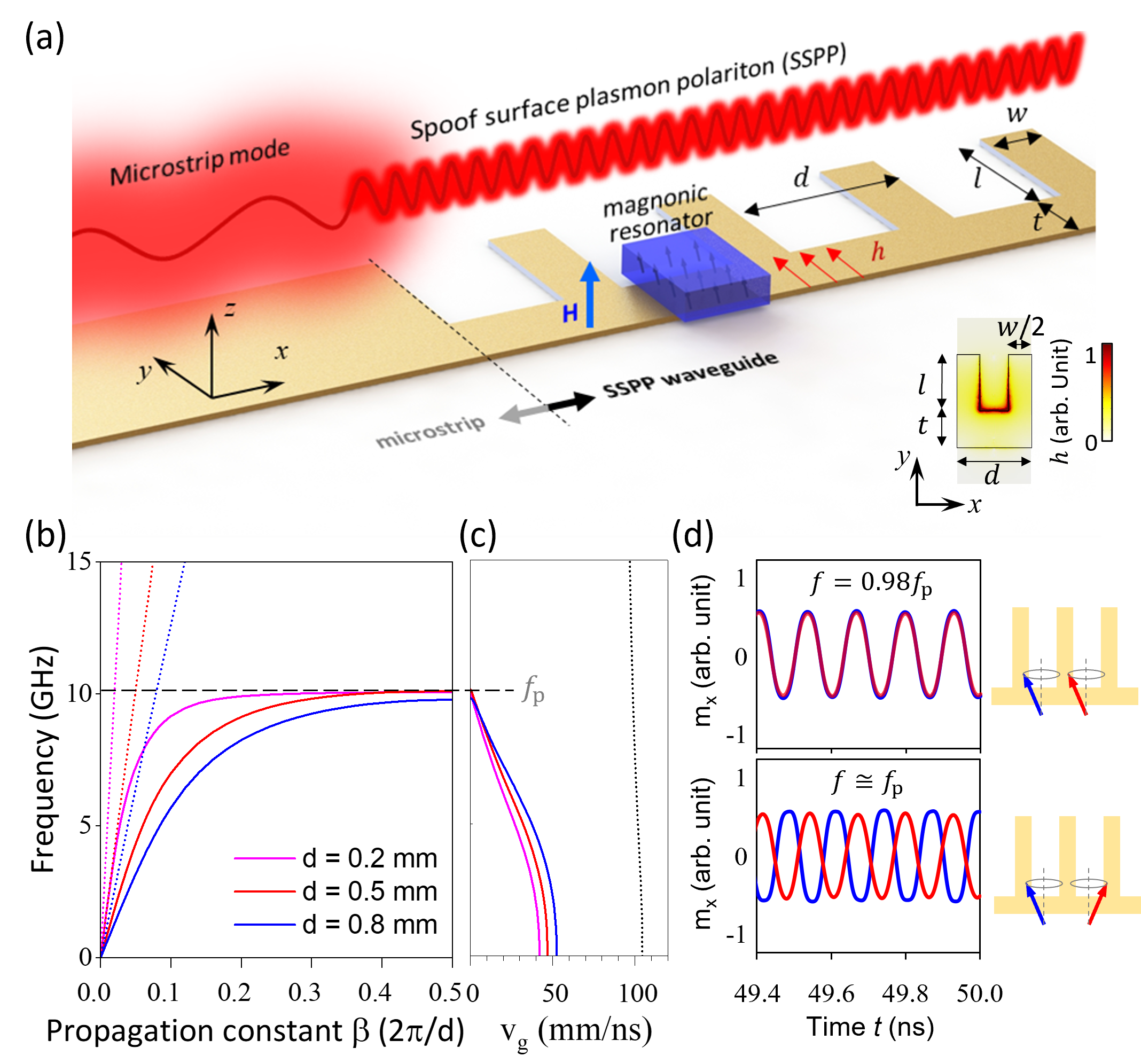}
\caption{(a) Schematics of the hybrid SSPP-magnonic device (not to scale). A planar magnonic resonator is placed on a periodically corrugated microstrip. Inset: simulated mode profile (total magnetic field $h$) of SSPPs. (b) Dispersions and (c) group velocities obtained from COMSOL simulation for SSPPs on a waveguide with $w=t=100 \mu$m, $l=3$ mm, and varying period $d$. Dotted lines: dispersions of the same uncorrugated microstrip, plotted against $\beta$ values that are normalized using different $d$ values. $f_p$: effective plasma frequency of SSPPs. (d) Phase-field simulation results for the temporal evolution of the precessing field $m_x$ of the magnon modes in two neighboring magnonic resonators (in phase at $f=0.98f_\mathrm{p}$; $\pi$ out of phase at $f\approx f_\mathrm{p}$).}
\label{fig1}
\end{figure}


Our slow-wave hybrid magnonic device consists of a conformal SSPP waveguide \cite{Shen_PNAS_2013} and a magnonic resonator [Fig.~\ref{fig1}(a)]. The SSPP waveguide is a metallic microstrip with periodic corrugations (period $d$), which supports slowly propagating SSPPs that mimic optical surface plasmons at metal/dielectric interfaces. Figure\,\ref{fig1}(b) plots the simulated dispersion curves of the fundamental SSPP modes on the waveguide (solid lines), which dramatically deviate from the dispersion of the uncorrugated microstrip modes (dotted lines) because of their polaritonic nature \cite{Gao_AdvMater_2018}. At the edge of the first Brillouin zone where the propagation constant $\beta = \pi/d$, the SSPP dispersion becomes nearly flat and asymptotically approaches the effective plasma frequency $f_\mathrm{p}$, where the group velocity is largely reduced [Fig.\,\ref{fig1}(c)].

The effective plasma frequency $f_\mathrm{p}$ of SSPPs is primarily determined by the corrugation depth $l$, which reaches around $10$ GHz when $l=3$ mm. The dispersion curve, and accordingly the group velocity, can be fine tuned by the corrugation period $d$ without significantly affecting $f_\mathrm{p}$ [Figs.\,\ref{fig1}(b) and (c)]. The corrugation teeth width $w$ has negligible effects and $w=0.1$ mm is used throughout our experiments. The dispersion curve and $f_\mathrm{p}$ are also sensitive to the magnonic resonator chip (size, alignment, bonding condition) and thus may vary from device to device in our experiment.

As the SSPP frequency approaches $f_\mathrm{p}$, the reduced group velocity is accompanied by the enhanced mode confinement [Fig.\,\ref{fig1}(a), inset], which can reach deep sub-wavelength level. The magnetic fields of SSPPs are strongly localized at the bottom of the corrugations, inducing enhanced coupling with magnons when planar magnonic resonators are placed there. The magnonic resonators are biased by an external magnetic field along $z$, which is perpendicular to the magnetic field (in $y$ direction) of SSPPs to ensure their proper coupling with magnons. The strength of the bias field, which determines the magnon frequency, is controlled by the $z$ position of the magnet using an automated stage.

The coherent SSPP-magnon coupling is confirmed by our phase-field simulation \cite{SM} based on coupled Maxwell's equations and Landau-Lifshitz-Gilbert equation \cite{Zhuang_NPJCM_2022,Zhuang_JPD_2023}. Our simulation shows that when the frequencies of SSPPs and magnons match, the oscillating magnetic field of SSPPs excites the precessing magnetization of magnon (and vice versa) through magnetic dipole-dipole interaction, and the magnon phase is determined by the phase of SSPP. Therefore, if multiple magnonic resonators are placed in series on an SSPP waveguide, their relative phase can be tuned by varying the operation frequency which accordingly changes the propagation constant. Thanks to the nearly flat dispersion near $f_p$, a small frequency tuning can produce a large phase tuning. This is verified by the simulated temporal evolution of $m_x$ [Fig.\,\ref{fig1}(d)], where the relative phase of two neighboring magnonic resonators (separated by $d$) varies from 0 to almost $\pi$ when the frequency is tuned by 2\%. Such a large phase tuning over a small distance is highly challenging when conventional waveguides such as microstrips or CPWs are used.
The considerable phase tunability, enabled by the slow-wave nature of SSPPs, presents a notable advantage over conventional resonator-based hybrid magnonic systems, where fixed phase detuning limits the system up-scaling for intricate functionalities or dynamics, such as programmable interference within a sizable array of magnonic resonators.

The coupling strength between magnons and traveling SSPPs can be calculated using Fermi's Golden rule
\begin{equation}
    g_\mathrm{ms} = 2\pi|g_0(\omega)|^2D(\omega)=A\frac{\hbar\omega}{v_\mathrm{g}S_\mathrm{eff}},
\label{couplingStrength}
\end{equation}
\noindent where $g_0(\omega)=\frac{\gamma}{2}\sqrt{\frac{2\hbar\omega\mu_0 s N}{V}}$ is the single magnon-SSPP coupling rate, $D(\omega) =\frac{L}{2\pi v_g}$ is the density of state of the SSPPs traveling in one direction, and the coefficient $A=\frac{1}{2}\eta^2\gamma^2\mu_0sN$. Here $\hbar$ is reduced Planck's constant, $\eta$ is the mode overlapping factor \cite{XufengZhang_PRL_2014}, $\gamma=28$ GHz/T is the gyromagnetic ratio, $\omega$ is the magnon angular frequency, $s$ is the spin quantum number, $\mu_0$ is the vacuum permeability, $N$ is the total number of spins, $V=S_\mathrm{eff}\times L$ is the effective interacting mode volume, $S_\mathrm{eff}$ is the effective cross-section area of the traveling SSPPs, $L$ is the length of the YIG resonator, and $v_\mathrm{g}=d\omega/d\beta$ is the group velocity of the traveling SSPPs. Compared with uncorrugated microstrips (width: $l+t$), SSPPs on our corrugated microstrip have significantly reduced mode cross-section area and group velocity \cite{SM}, which will therefore lead to drastically increased coupling strengths between magnons and the microwave photons in the SSPP mode.

\begin{figure}[bt]
\centering
\includegraphics[width=0.98\linewidth]{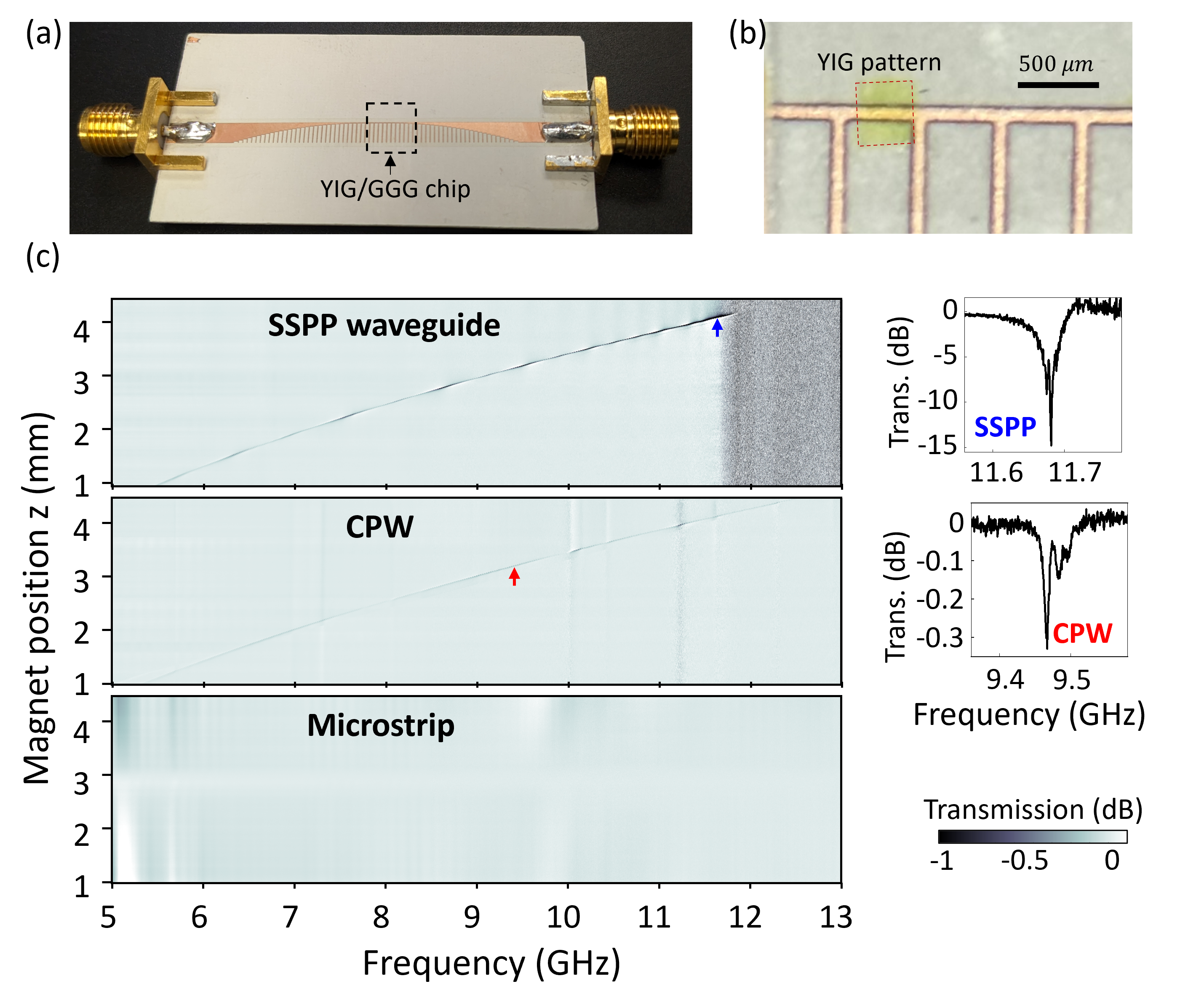}
\caption{(a) Optical image of the fabricated SSPP waveguide circuit. Between the SSPP waveguide and the SMA connectors there exists a tapered transition region where the corrugation depth $l$ is quadratically varied. (b) Zoomed-in image showing a YIG resonator ($400 \mu m \times 400 \mu m$) flipped on the SSPP waveguide. (c) Measured transmission spectra for SSPP waveguide, CPW, and microstrip, respectively. Background spectra are removed to highlight the weak magnon resonances. Insets: the measured spectra corresponding to the points indicated by the arrows, respectively.}
\label{fig2}
\end{figure}

In our experiment, an SSPP waveguide with a 500 $\mu m$ corrugation period is fabricated on a high-dielectric constant substrate ($\varepsilon=9.8$) [Fig.\,\ref{fig2} (a)]. Magnonic resonators (lateral sizes range from tens to hundreds of micrometers) are fabricated on a 200-nm thin film of ferrimagnetic insulator yttrium iron garnet (YIG) epitaxially grown on a 500-$\mu$m gadolinium gallium garnet (GGG) substrate. The magnonic chip is flip-bonded to the SSPP waveguide circuit [Fig.\,\ref{fig2} (b)]. An out-of-plane bias magnetic field is applied using a permanent magnet, which can be moved along $z$ direction to control the magnon frequency.

The transmission signals measured using a vector network analyzer on a device having a 400 $\mu m \times$ 400 $\mu m\times$ 200 nm YIG resonator is plotted in Fig.\,\ref{fig2} (c). As a comparison, transmissions from a CPW (width of center line: 100 $\mu$m, gap to the ground: 500 $\mu$m) and a uncorrugated microstrip (width: 3 mm) loaded with the same YIG resonator are also plotted. In all spectra, the transmission background is removed to reveal the small magnon absorption dips, which are observed in a broad frequency range ($>7$ GHz). A maximum extinction ratio of 15 dB is obtained at 11.68 GHz (near $f_\mathrm{p}$), whereas on the CPW it is two orders of magnitude smaller (0.3 dB at 9.46 GHz for the same magnon mode). On the uncorrugated microstrip, magnon modes are not observed. These results indicate the drastically enhanced SSPP-magnon  interaction due to the small group velocity of SSPPs, which is highly advantageous over CPWs/microstrips particularly for miniaturized magnonic devices.

\begin{figure}[tb]
\includegraphics[width=0.98\linewidth]{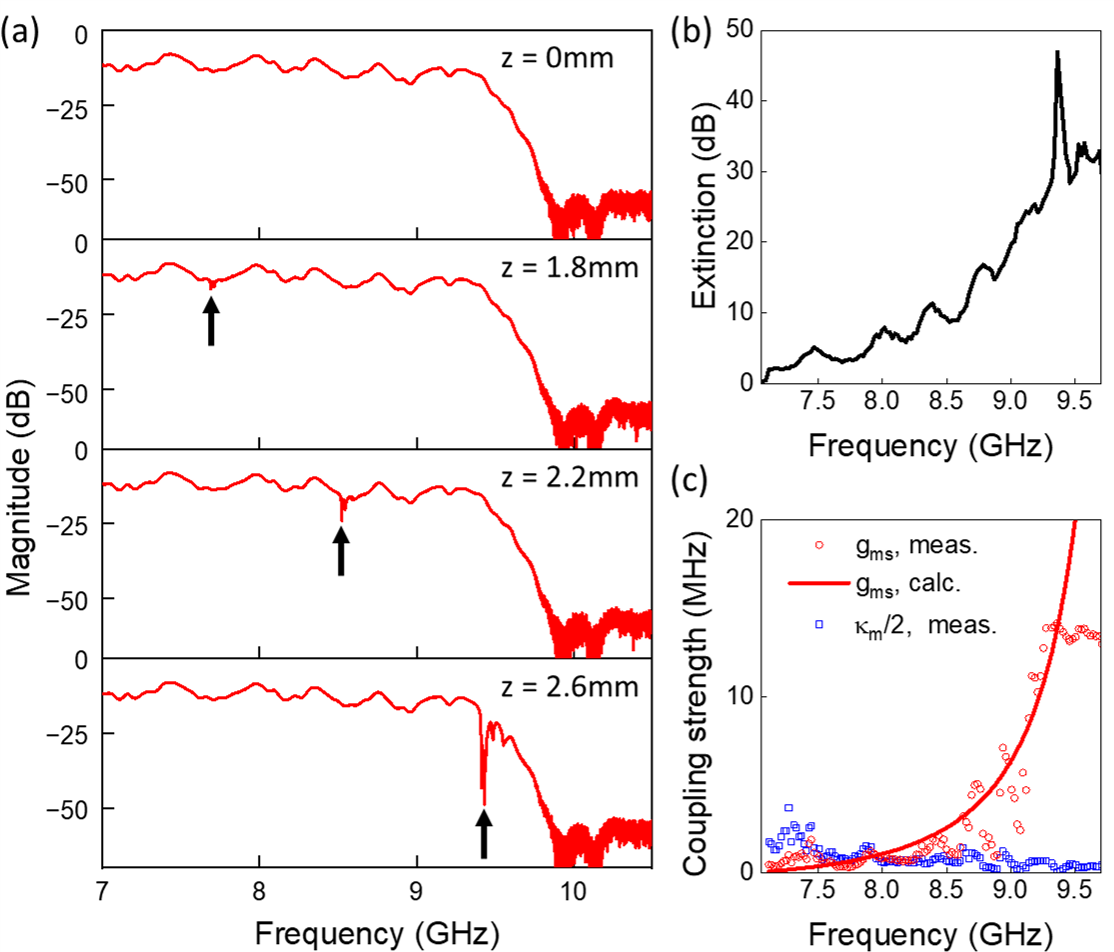}
\caption{(a) Measured transmission spectra of the SSPP waveguide with the magnet at different positions $z$. Black arrows indicate the magnon resonances. (b) Extracted extinction ratio of the magnon resonances as a function of the magnet frequency. (c) Extracted (red circles) and calculated (solid line) SSPP-magnon coupling strength $g_\mathrm{ms}$ as a function of magnon frequency. Blue squares: extracted intrinsic magnon damping rate $\kappa_m/2$.}
\label{fig3}
\end{figure}

The slow-wave enhancement of the SSPP-magnon coupling has a strong frequency dependence, as shown by the transmission spectra [Fig.\,\ref{fig3}(a)] for different magnon frequencies (determined by the magnet position $z$) on a device with a 300 $\mu $m $\times$ 1000 $\mu $m $\times$ 200 nm YIG resonator. When the magnet is at $z=0$ mm, the magnon mode is absent and the spectrum shows the intrinsic characteristics of SSPPs: a broad transmission band with a cutoff at around 10 GHz ($\approx f_\mathrm{p}$). Below the cutoff frequency, an insertion loss of about 12 dB is measured, which is attributed to the metal absorption and coupling loss with the RF connectors, while above the cutoff frequency, the transmission drops to below -50 dB because SSPPs are no longer supported. When the bias magnet moves towards the device, magnon modes are observed over a broad frequency range as narrow absorption dips in the transmission spectrum. For instance, when the magnet is at $z=1.8$ mm, magnon resonances are visible at 7.7 GHz, which further increases to 8.5 GHz at $z=2.2$ mm and 9.4 GHz at $z=2.6$ mm, respectively. 

Figure\,\ref{fig3}(a) reveals one striking feature of these magnon resonances: their extinction ratio increases as the magnon frequency approaches the SSPP cutoff frequency. Such dependence is clearly shown by the extracted extinction ratio as a function of the magnon frequency [Fig.\,\ref{fig3} (b)]. The small oscillations in the curve are due to the interference effects of the SSPPs when propagating along the waveguide with a finite length, but they are much weaker than the enhancement effect caused by the increased magnon frequency. A maximum extinction ratio of 47 dB is observed at the edge of the cutoff frequency ($z=2.6$ mm), which is more than four orders of magnitude higher than what is obtained near $z=1.8$ mm.

The SSPP-magnon coupling strength [Fig.\,\ref{fig3}(c), red circles] is extracted from the measured spectra \cite{SM}. Despite the fluctuation caused by the background interference (8.5 - 9 GHz) \cite{SM}, an increasing trend with magnon frequency is evident. Because $S_\mathrm{eff}\propto 1/\beta^2$ \cite{SM}, Eq.\,[\ref{couplingStrength}] can be rewritten as $g_\mathrm{ms}=A'\omega \beta^2/v_g$, where the phenomenological coefficient $A'$ is the only unknown parameter since both the dispersion $\beta(\omega)$ and group velocity $v_g(\omega)$ can be obtained from simulations. Good agreement is obtained between our theoretical calculation [solid line in Fig.\,\ref{fig3} (c)] and the measurement results, with an exception above 9.3 GHz where the reduced transmission causes inaccurate numerical fittings. A maximum SSPP-magnon coupling strength of nearly 15 MHz are measured at 9.2 GHz, which is 20 times larger than the extracted intrinsic magnon damping rate $\kappa_m/2=0.7$ MHz. Even larger coupling strengths are expected, but their observation is hindered by the increased SSPP loss (accordingly, transmission drop) near $f_p$.

\begin{figure}[tb]
\includegraphics[width=\linewidth]{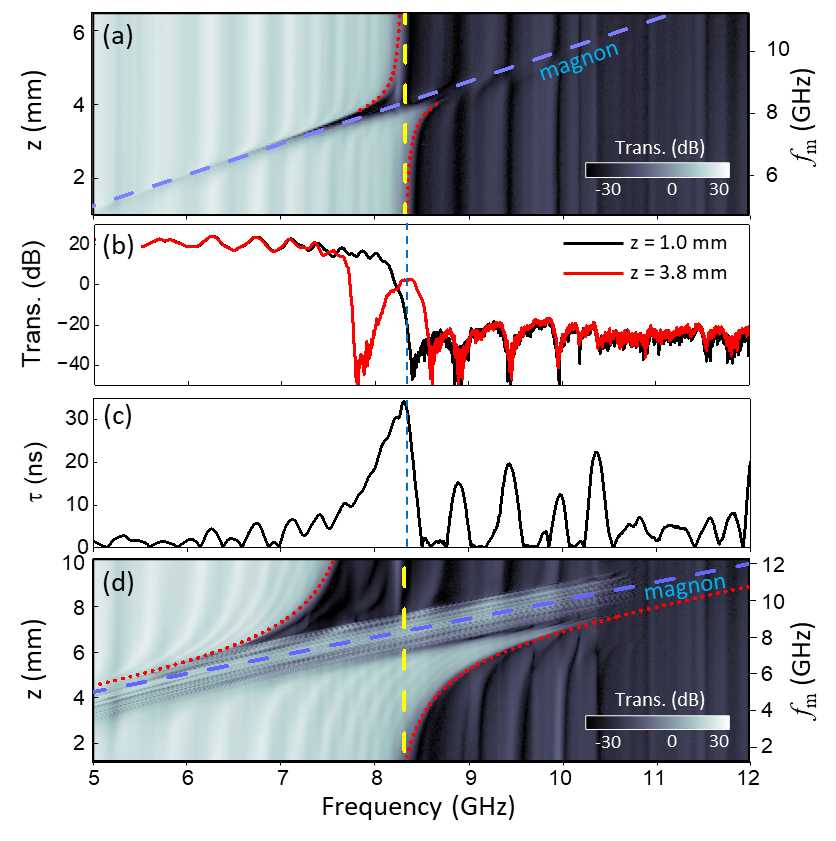}
\caption{(a) Measured transmission spectra of a SSPP waveguide covered by 3 $\mu$m $\times$ 5 mm $\times$ 5 mm rectangular YIG resonator. (b) The measured SSPP waveguide transmission at magnet position $z=1.0$ mm (black) and $z=3.8$ mm (red), respectively. (c) Extracted group delay $\tau$ from the measured transmission phase. The large oscillations above the cutoff frequency are due to standing waves.  (d) The measured transmission spectra using a 350-$\mu$m-thick, 5-mm-diameter YIG disc. Trans.: transmission. All dashed and dotted lines in (a) and (d) are from calculation which uses right $y$ axis. Blue line: calculated magnon mode; red and yellow lines: calculated $f_\mathrm{p}$ with and without SSPP-magnon coupling, respectively. $f_m$ is the magnon frequency. }
\label{fig4}
\end{figure}

Similar to cavity hybrid magnonics, SSPP-magnon coupling can be enhanced by increasing the YIG volume. Figure\,\ref{fig4} plots the measured transmission for a 3 $\mu$m $\times$ 5 mm $\times$ 5 mm YIG resonator on the same SSPP waveguide. An anti-crossing feature with a splitting $2g_m=0.7$ GHz is observed near the cutoff frequency around 8.4 GHz [Fig.\,\ref{fig4}(b)]. This is modeled based on the frequency response of YIG's permeability near the magnonic resonance and its effect on the effective plasma frequency of SSPPs. The theoretical prediction shows great agreement with the measurement results [Fig.\,\ref{fig4}(a)], with all four curves computed concurrently using a single equation involving only two fitting parameters \cite{SM}. The coupling strength $g_m=0.35$ GHz corresponds to an oscillation period $1/g_m=2.9$ ns for the coherent information exchange between magnons and SSPPs. On the other hand, the measured group delay $\tau=d \phi / d \omega$ of the whole 35-mm-long SSPP waveguide reaches a maximum of 32 ns [Fig.\,\ref{fig4}(c)], yielding a traveling time $\tau_g=4.6$ ns for the SSPPs to pass the YIG resonator. Since the traveling time exceeds the oscillation period, the coherent signal will experience multiple oscillations between the magnon and SSPP modes while traveling through the interaction region, which can be defined as the slow-wave strong coupling condition: $\tau_g > 1/g_\mathrm{ms}$. By replacing the 3-$\mu$m-thick YIG resonator with a thick YIG disc (thickness: 350 $\mu$m, diameter: 5 mm), the slow-wave SSPP-magnon strong coupling can be further enhanced with a splitting of 3.2 GHz ($g_m=1.6$ GHz) as obtained through our modeling [Fig.\,\ref{fig4}(d)], leading to a single-spin coupling strength $g_0=130$ Hz. Considering that $g_m/f_\mathrm{p}=19.3\%$, this corresponds to the ultrastrong coupling in conventional hybrid magnonics.

This work demonstrates a novel type of slow-wave hybrid magnonics based on SSPP-magnon interaction. Thanks to the small group velocities of the SSPPs, our device platform exhibits significantly enhanced coupling strength while maintaining a large operation bandwidth, combining the advantages of cavity magnonics and traveling wave devices. Our slow-wave hybrid magnonic system opens a new chapter for hybrid magnonics, promising potential applications in integrated magnonic circuits \cite{Wang_sciadv_2018Jan,Wang_natelectro_2020Dec} and quantum information processing\cite{Tabuchi_science_2015Jul,Lachance-Quirion_APE_2019Jun,Lachance-Quirion_Science_2020Jan,Awschalom2021Feb,Hirosawa_PRX_2022Nov,Yuan_PR_2022Jun,Xu_PRL_2023May}. 
Moreover, our demonstrated principle can be extended to other hybrid magnonics systems such as optomagnonics and magnomechanics. This work also points to a new direction for the study and application of spoof plasmonics.

\begin{acknowledgments}
The authors thank R. Divan, L. Stan, and C. Miller for supports in the device fabrication. X.Z. acknowledges support from ONR YIP (N00014-23-1-2144). C.Z. and L.J. acknowledge support from the ARO (W911NF-23-1-0077), AFRL (FA8649-21-P-0781), NSF (OMA-2137642), NTT Research, and Packard Foundation (2020-71479). L.J. acknowledges the support from the Marshall and Arlene Bennett Family Research Program. J.H. acknowledges support from the National Science Foundation (NSF) Award No. CBET-2006028. The works on dynamical phase-field simulations used Bridges2-GPU at Pittsburgh Supercomputing Center through allocation DMR180076 from the Advanced Cyberinfrastructure Coordination Ecosystem: Services \& Support (ACCESS) program, which is supported by National Science Foundation (grant numbers 2138259, 2138286, 2138307, 2137603, and 2138296. J.M.J. acknowledges funding support from the US National Science Foundation (grant number 2011411). Work performed at the Center for Nanoscale Materials, a U.S. Department of Energy Office of Science User Facility, was supported by the U.S. DOE, Office of Basic Energy Sciences, under Contract No. DE-AC02-06CH11357. 
\end{acknowledgments}

\bibliography{Maintext}

\end{document}


\preprint{APS/123-QED}

\title{Supplemental Material for Slow-Wave Hybrid Magnonics}

\author{Jing Xu}
\affiliation{ 
    Center for Nanoscale Materials, Argonne National Laboratory, Lemont, IL 60439, USA
}

\author{Changchun Zhong}
\affiliation{ 
    Pritzker School of Molecular Engineering, University of Chicago, Chicago, IL 60637, USA
}

\author{Shihao Zhuang}
\affiliation{ 
    Department of Materials Science and Engineering, University of Wisconsin - Madison, Madison, WI 53706, USA
}

\author{Chen Qian}
\affiliation{ 
    Department of Physics and Astronomy, 
    University of Pennsylvania, Philadelphia, PA, 19104, USA
}

\author{Yu Jiang}
\affiliation{ 
    Department of Electrical and Computer Engineering, Northeastern University, Boston, MA 02115, USA
}

\author{Amin Pishehvar}
\affiliation{ 
    Department of Electrical and Computer Engineering, Northeastern University, Boston, MA 02115, USA
}

\author{Xu Han}
\affiliation{ 
    Center for Nanoscale Materials, Argonne National Laboratory, Lemont, IL 60439, USA
}

\author{Dafei Jin}
\affiliation{
    Department of Physics and Astronomy, University of Notre Dame, Notre Dame, Indiana 46556, USA
}
\affiliation{ 
    Center for Nanoscale Materials, Argonne National Laboratory, Lemont, IL 60439, USA
}

\author{Josep M. Jornet}
\affiliation{ 
    Department of Electrical and Computer Engineering, Northeastern University, Boston, MA 02115, USA
}

\author{Bo Zhen}
\affiliation{ 
    Department of Physics and Astronomy, 
    University of Pennsylvania, Philadelphia, PA, 19104, USA
}

\author{Jiamian Hu}
\affiliation{ 
    Department of Materials Science and Engineering, University of Wisconsin - Madison, Madison, WI 53706, USA
}

\author{Liang Jiang}
\affiliation{ 
    Pritzker School of Molecular Engineering, University of Chicago, Chicago, IL 60637, USA
}

\author{Xufeng Zhang}
\email{xu.zhang@northeastern.edu}
\affiliation{ 
    Department of Electrical and Computer Engineering, Northeastern University, Boston, MA 02115, USA
}

\date{\today}


\maketitle



\section{Slow-wave-enhanced coupling strength}

Denoting the traveling wave in the waveguide as $q(t,x)$, it is expressed as $q(t,x)=\frac{1}{\sqrt{2\pi}}\int d\omega q(\omega) e^{-i\omega(t-x/v_g)}$, where $v_g$ is the group velocity and $q(\omega)$ is the bosonic operator satisfying the normalization $[q(\omega),q^\dagger(\omega^\prime)]=\delta(\omega-\omega^\prime)$. The interaction between the magnon and the wave guide is through the Hamiltonian 
\begin{equation} 
H_I=\frac{i\hbar}{\sqrt{2\pi}}\int d\omega f_q(\omega)*(\hat{m}q^\dagger(\omega)+\hat{m}^\dagger q(\omega)).
\end{equation}
The total coupling rate of the magnon to the wave guide can be estimated by the Fermi Golden rule
\begin{equation}
g_\mathrm{ms}=2\pi |f_q(\omega)|^2D(\omega).
\end{equation}
$f_q(\omega)$ takes the form $f_{q}(\omega)=\frac{\gamma}{2}\sqrt{\frac{\hbar\omega\mu sN}{V}}$,
where $V$ is the effective mode volume, $N$ is the total number of spin and $s$ is the spin number, respectively. The effective volume can be expressed $V=S_\text{eff}\times L$, where $S_\text{eff}$ is the effective mode cross section of the traveling wave and $L$ is the YIG length. The photon density of state traveling in one direction is given as $D(\omega)d\omega=\frac{L}{2\pi v_g}d\omega.$ Putting them together, we have 
\begin{equation}
    g_\text{ms}=2\pi \gamma^2\frac{1}{4}{\frac{\hbar\omega\mu sN}{2\pi}} \frac{1}{S_\text{eff} v_g}.
\end{equation}
Obviously, the coupling strength can be greatly enhanced if we can reduce the traveling wave group velocity $v_g$ and the effective cross section $S_\text{eff}$. Note that the SSPP lives on the material boundary, and effectively the field's cross section depends on the field penetration depth. The penetration depth into the two materials can be estimated as $l_1={1}/{\sqrt{\beta^2-k_0^2}}$, $l_2={1}/{\sqrt{\beta^2-k_0^2+k_p^2}}$, where $\beta$ is the propagation constant, $k_0$ is the wave vector in free space and $k_p$ is the plasma wave vector. Since the effective cross section is proportional to the penetration length $S_\text{eff}\propto (l_1+l_2)^2$, we obtain the coupling strength
\begin{equation}
\begin{split}
    g_\text{ms}&\propto\frac{\omega}{v_g\times (l_1+l_2)^2}\\
    &=\frac{\omega}{v_g}\times\left(\frac{1}{\sqrt{{\beta^2}-{k_0^2}}}+\frac{1}{\sqrt{{\beta^2}-{k_0^2}+{k_p^2}}}\right)^{-2},
\end{split}   
\end{equation}
showing an enhanced coupling rate as the system approaching the low group velocity regime. When $\beta$ is large, this expression reduces to
\begin{equation}
    g_\text{ms}\propto \frac{\omega}{v_g}\beta^2.
\end{equation}

\section{Coupling strength extraction}

\begin{figure}[htb]
\includegraphics[width=0.5\linewidth]{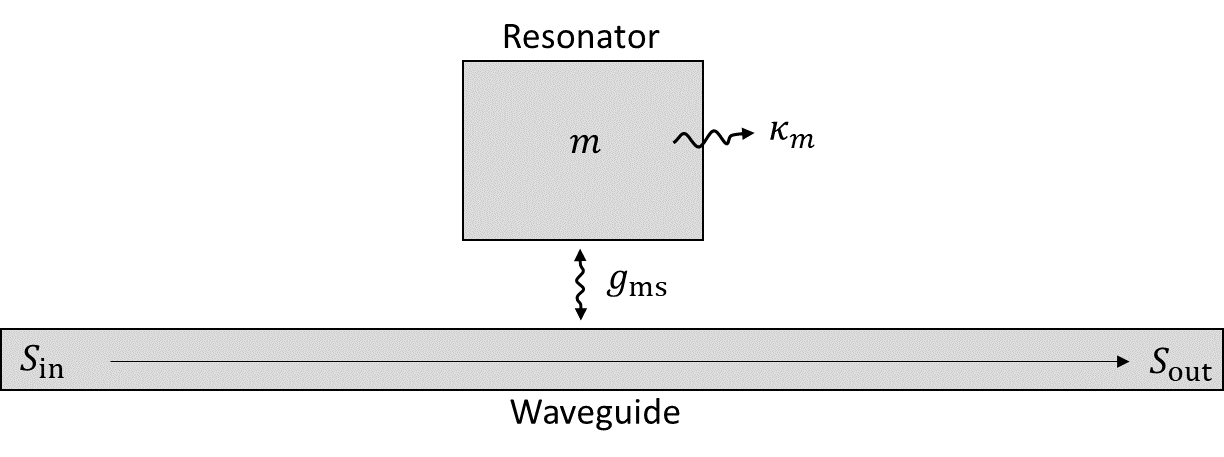}
\caption{Simplified model of a resonator coupled to a waveguide.}
\label{Resonator_Trans}
\end{figure}

Figure\,\ref{Resonator_Trans} shows the simplified model of our system that can be used to describe the coupling of the magnonic resonator with the SSPP waveguide. The input and output amplitudes of the travelling SSPP mode is $S_\mathrm{in}$ and $S_\mathrm{out}$, respectively. The coupling strength between the magnon mode and the SSPP mode is $g_\mathrm{ms}$, and the intrinsic magnon dissipation rate is $\kappa_m/2$.

The Equation of motion for the intra-cavity magnon mode $m$ is:
\begin{equation}
    \frac{dm}{dt} = i \Delta m-\frac{\kappa_t}{2} m + S_\mathrm{in}\sqrt{g_\mathrm{ms}},
\end{equation}
\noindent where $\Delta = \omega_m - \omega$ is the frequency detuning, $\omega_m$ is the magnon resonance frequency, $\omega$ is the input microwave frequency, $\kappa_t=\kappa_m+2g_\mathrm{ms}$ is the total magnon dissipation rate, $\kappa_m$ is the intrinsic magnon damping rate, $g_\mathrm{ms}$ is the magnon-SSPP coupling rate, $S_\mathrm{in}$ is the input amplitude, respectively.\\

At steady state, we have $\frac{dm}{dt}=0$, and accordingly
\begin{equation}
    (i\Delta + \frac{\kappa_t}{2}) m =S_\mathrm{in} \sqrt{g_\mathrm{ms}},
\end{equation}
\noindent which gives the intra-cavity magnon as
\begin{equation}
    m = \frac{S_\mathrm{in}\sqrt{g_\mathrm{ms}}}{i\Delta + \frac{\kappa_t}{2}}.
\end{equation}

Consequently, the output SSPP amplitude can be calculated as 
\begin{align*}
    S_\mathrm{out} & = S_\mathrm{in}-m\sqrt{g_\mathrm{ms}}  \\
    & = S_\mathrm{in}\left(1-\frac{g_\mathrm{ms}}{i\Delta+\frac{\kappa_t}{2}}\right).
\end{align*}

Therefore, the transmission can be obtained as 
\begin{equation}
    T=\left | \frac{S_\mathrm{out}}{S_\mathrm{in}} \right |^2=\left | 1-\frac{g_\mathrm{ms}}{i\Delta+\frac{\kappa_t}{2}}\right | ^2 = \frac{4\Delta^2+\kappa_m^2}{4\Delta ^2 + \kappa_t^2}
\end{equation}.

When on resonance ($\Delta=0$), the transmission becomes
\begin{equation}
    T=\left (\frac{\kappa_m}{\kappa_t}\right )^2=\left(1-\frac{2g_\mathrm{ms}}{\kappa_t}\right ) ^2.
\end{equation}

Therefore, the coupling rate between the magnon and SSPP mode can be determined as
\begin{equation}
    g_\mathrm{ms}=\frac{1-\sqrt{T}}{2}\kappa_t,
\end{equation}
\noindent Both the on resonance transmission $T$ and the total magnon dissipation rate can be extracted from the measurement results.

\section{Modeling of SSPP-magnon Strong coupling}
The effective plasma frequency $f_\mathrm{p}$ of the SSPP mode is determined by the resonance frequency of a quarter-wave resonator which has a length equal the corrugation depth ($l$) of the SSPP waveguide. Therefore, $f_\mathrm{p}$ is directly influenced by the permittivity and permeability of the surrounding environment. In our device, the bottom cladding of the SSPP waveguide is a dielectric with a relative permittivity $\varepsilon_\mathrm{die}=9.8$ and relative permeability of $\mu_\mathrm{die}=1$, and the top cladding is the YIG chip, which has a relative permittivity $\varepsilon_\mathrm{YIG}=15$ and an effective relative permeability $\mu_\mathrm{YIG}=\mu'-\frac{\mu''^2}{\mu'}$, where $\mu'$ and $\mu''$ are the diagonal and off-diagonal components of $\mu_\mathrm{YIG}$, respectively. For a normally biased YIG film with a field strength $H_\mathrm{ext}$ and saturation magnetization $M_s$, we have $\mu'=1+\frac{f_mf_0}{f_0^2-f^2}$, and $\mu''=\frac{f_mf}{f_0^2-f^2}$, where $f_m=\gamma \times 4\pi M_s$, $f_0=\gamma (H_\mathrm{ext}-4\pi M_s)$, $\gamma=28$ GHz/T is the gyromagnetic ratio, and $f$ is the operation frequency.

The resonance frequency of the quarter-wave resonator with a length $l$ can be calcuated as $f=\frac{c}{n\lambda}=\frac{c}{4nl}$, where $c$ is the speed of light in vacuum and $n$ is the effective refractive index which can be approximated by the average refractive indices of both the top and bottom claddings $n=\frac{1}{2}(n_\mathrm{die}+n_\mathrm{YIG})=\frac{1}{2}(\sqrt{\varepsilon_\mathrm{die}\mu_\mathrm{die}}+\sqrt{\varepsilon_\mathrm{YIG}\mu_\mathrm{YIG}})$. Therefore, the resonance frequency can be calculated as 

\begin{equation}
    f_\mathrm{p}=\frac{c}{2l(\sqrt{\varepsilon_\mathrm{die}\mu_\mathrm{die}}+\sqrt{\varepsilon_\mathrm{YIG}\mu_\mathrm{YIG}})}.
\end{equation}

The above equation describes the ideal situation. In our device, we have to consider that fact that the SSPP waveguide is not fully covered by the YIG chip but instead partially exposed to air, and thus the average refractive index should be $n=\frac{1}{2}(n_\mathrm{die}+n_\mathrm{YIG}')=\frac{1}{2}[n_\mathrm{die}+(1-r_1)n_\mathrm{air}+r_1 n_\mathrm{YIG}]$, where $n_\mathrm{air}=\sqrt{\varepsilon_\mathrm{air}}=1$ is the refractive index of air. Here $r_1$ is the mixing ratio of YIG and air in the top cladding, which is a fitting parameter in our calculation. \\

In addition, we also have to consider the fact that not all the spins in YIG are equally contributing to the SSPP-magnon coupling, considering the nonuniform distribution (both intensity and direction) of the magnetic field of the SSPP mode. Therefore, the permeability of the top cladding should be averaged between YIG and a nonmagnetic medium with a permeability of 1: $\mu_\mathrm{YIG}'=[(1-r_2)\times 1 + r_2\mu_\mathrm{YIG}]$.\\

\begin{figure}[htb]
\includegraphics[width=0.9\linewidth]{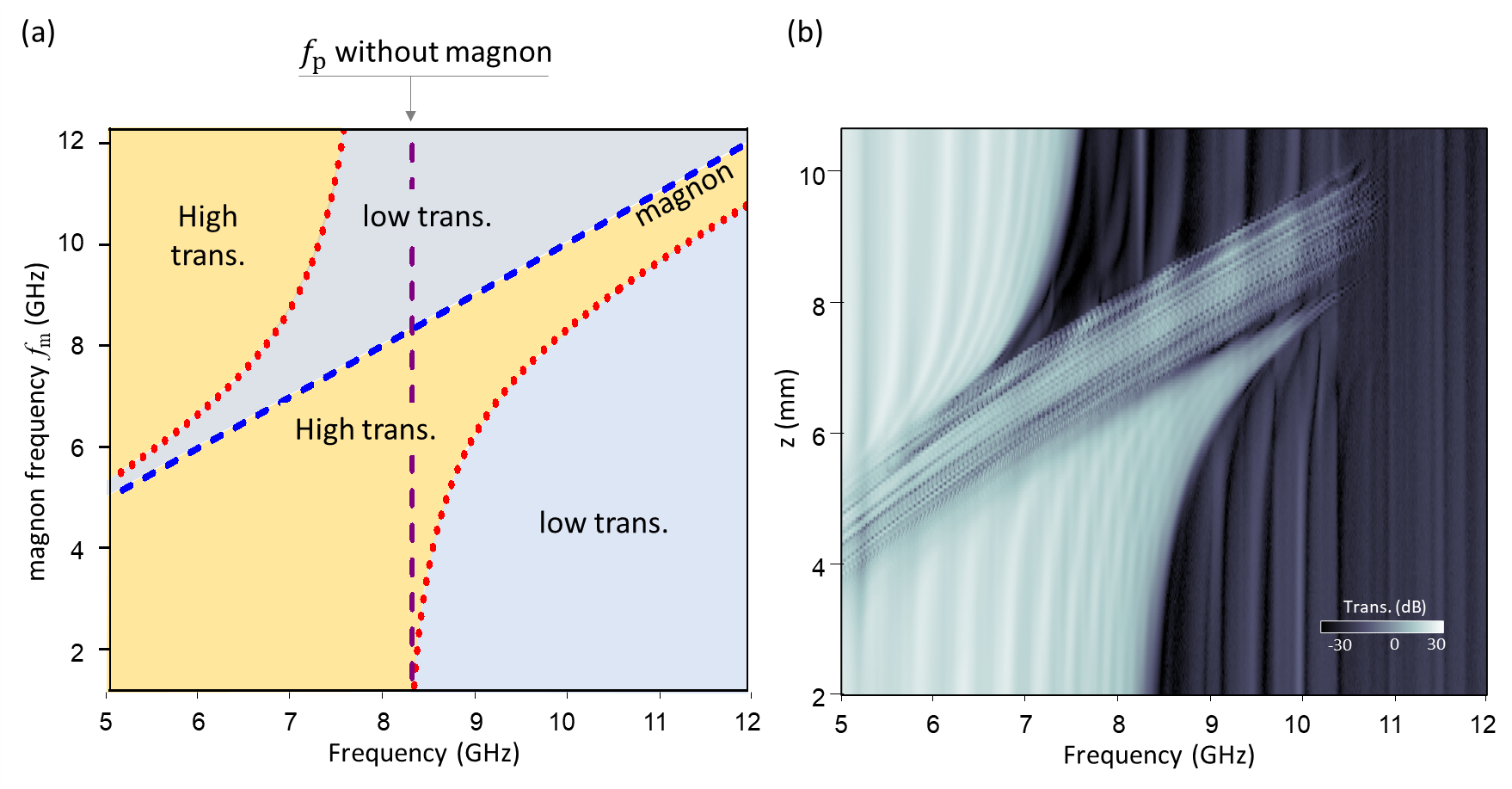}
\caption{(a) Calculated spectrum using Equation\,\ref{Eq:fp}. (b) The same measured transmission spectrum as plotted in Fig. 4(d) of the main text.}
\label{fig_SM_calSpectrum}
\end{figure}

Consequently, the resonance frequency now becomes
\begin{equation}
    f_\mathrm{p}=\frac{c}{2l}\cdot \frac{1}{n_\mathrm{die}+(1-r_1)n_\mathrm{air}+r_1n_\mathrm{YIG}}=\frac{c}{2l}\cdot\frac{1}{\sqrt{\varepsilon_\mathrm{die}\mu_\mathrm{die}}+(1-r_1)\sqrt{\varepsilon_\mathrm{air}}+r_1\sqrt{\varepsilon_\mathrm{YIG}(1-r_2+r_2\mu_\mathrm{YIG})}}.
\end{equation}
\noindent Taking into account the magnetic response of YIG's permeability, the resonance frequency can be expressed as
\begin{equation}
    f_\mathrm{p}=\frac{c}{2l}\cdot \frac{1}{\sqrt{\varepsilon_\mathrm{die}}+(1-r_1)+r_1\sqrt{\varepsilon_\mathrm{YIG}\left(1-r_2+r_2\frac{f_\mathrm{c}^2-f_\mathrm{p}^2}{f_{\mathrm{FMR}}^2-f_\mathrm{p}^2}\right)}},
    \label{Eq:fp}
\end{equation}
\noindent where $f_\mathrm{FMR}=\sqrt{f_0^2+f_0f_m}$ is the ferromagnetic resonance frequency for a normally magnetized YIG film, and $f_c=f_0+f_m$. Note that in this equation, the effective plasma frequency $f_\mathrm{p}$ appears on both sides of the equation, and thus the solution of the equation might not be single valued. Our calculation results show that near the magnon resonance frequency, $f_\mathrm{p}$ split into two values, as shown in Fig.\,\ref{fig_SM_calSpectrum} and Fig. 4 (a) \& (d) of the main text. It is worth noting that all the four curves, i.e., the intrinsic SSPP plasma frequency (when magnon is absent), the magnon frequency, and the two split SSPP plasma frequency curves (due to SSPP-magnon coupling), are obtained concurrently using the same Equation\,\ref{Eq:fp}, and the only two unknown fitting parameters in the calculation are the two mixing ratios $r_1$ and $r_2$. Figures\, 4(a)\&(d) in the main text and Fig.\,\ref{fig_SM_calSpectrum} clearly show that our calculation results reproduced the measurement results very well, validating our theoretical model. Based on these calculations, a splitting of $0.7$ GHz and $3.2$ GHz are obtained for Fig.4(a) and Fig.4(b) of the main text, respectively. The splitting is determined as the difference of the two $f_\mathrm{p}$ values measured at the frequency where the magnon and intrinsic SSPP plasma frequencies intersect.\\

Since the $f_\mathrm{p}$ represents the edge of the SSPP band, high (low) transmission is expected below (above) each $f_\mathrm{p}$. Specifically, high transmission is expected when the inequality
\begin{equation}
    f<\frac{c}{2l}\cdot \frac{1}{\sqrt{\varepsilon_\mathrm{die}}+(1-r_1)+r_1\sqrt{\varepsilon_\mathrm{YIG}\left(1-r_2+r_2\frac{f_\mathrm{c}^2-f^2}{f_{\mathrm{FMR}}^2-f^2}\right)}},
    \label{Eq:fp}
\end{equation}
\noexpand is satisfied, where $f$ is the measurement frequency, and when the inquality is unsatisfied, low transmissions are expected. The high transmission regions are highlighted in yellow in the calculated spectrum in Fig.\,\ref{fig_SM_calSpectrum}, whereas the low transmission regions are highlighted in gray.

\section{GPU-accelerated dynamical phase-field simulations of SSPP-magnon coupling}

\begin{figure}[htb]
\includegraphics[width=0.6\linewidth]{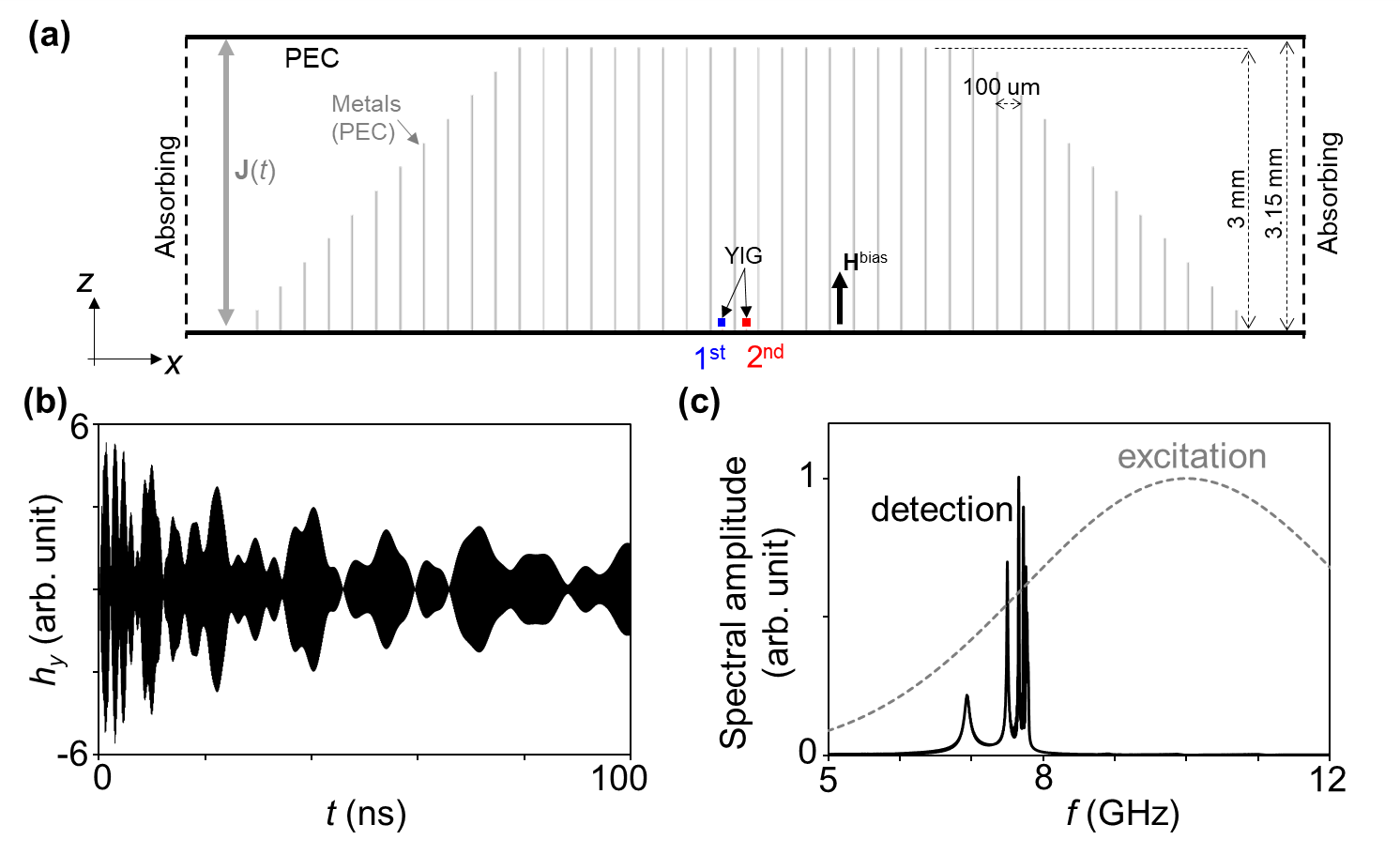}
\caption{(a) Illustration of the SSPP waveguide structure used in the simulations. (b) Evolution of magnetic-field component hy of the EM wave at the position of the ‘2nd YIG` under bias magnetic field $H^\mathrm{bias}$ = 0. (c) Frequency spectra of the excitation charge pulse (dashed line) and the detected field signal (solid line).}
\label{fig_SM_config}
\end{figure}

Using our in-house dynamical phase-field model that incorporates the coupled dynamics of magnons and photons in multiphase system \cite{SM_Zhuang_NPJCM_2022}, we model the SSPP-magnon coupling in the waveguide in a system as illustrated in Fig.\,\ref{fig_SM_config}(a), where all physical quantities are assumed to be uniform along the $y$ axis. To speed up the simulation and suppress numerical errors, a simplified two-dimensional (2D) model is used in our simulation. Previous studies have shown that the conformal SSPPs are insensitive to the SSPP waveguide thickness \cite{Shen_PNAS_2013} and therefore the behavior of the 2D model (i.e., infinite metal thickness) well represents the 3D model (metal thickness $\sim 20 \mu$m) used in our experiments. Two YIG resonators are placed in two neighboring grooves (between the metallic teeth). A time-dependent charge current $\textbf{J}(t)$ is applied along the $z$ axis [see Fig.\,\ref{fig_SM_config}(a)] to excite the SSPP mode and the magnetization precession. The magnetization orientation is initially aligned the $+z$ axis, stabilized by the bias magnetic field $\textbf{H}^\mathrm{bias}$ applied along the same direction. Note that in this 2D model, the microwave magnetic field is along the out-of-plane direction ($y$ direction in this case), and therefore to ensure nonzero magnon-SSPP coupling, the bias field can only be applied along $z$ direction (in-plane). This is different from the 3D configuration in Fig. 1(a) where the bias field is applied out-of-plane, because in that case the microwave magnetic components have both in-plane and out-of-plane components, so a bias field along either $y$ or $z$ will lead to nonzero coupling with the magnon mode. In fact, the specific bias field direction does not affect the operation principle of the enhanced coupling between SSPP and magnon modes, as long as the bias field is not parallel to the microwave magnetic fields which will lead to zero coupling. The ferromagnetic resonance (FMR) frequency (frequency of the Kittel mode magnon) can be tuned by varying the amplitude of  $\textbf{H}^\mathrm{bias}$. Such a simulation setup can well capture the fundamental physical principles of SSPPs and their interaction with magnons. More importantly, our time-domain simulations can provide critical information (both the amplitude and phase) about the temporal evolution of the SSPP and magnon excitations. The implementation of graphics processing unit (GPU) parallelization in our in-house codes enables a high-throughput simulation of the dependence of the dynamical SSPP-magnon coupling on key materials parameters such as magnetic damping coefficient and saturation magnetization (i.e., the spin density in the magnon resonator), which not only benefits this work but also allows future exploration of different magnonic materials.

In the phase-field simulations, the time-domain dynamics of the electromagnetic (EM) wave in the entire system and the magnetization in YIGs are simulated by solving the coupled Maxwell’s equations and Landau-Lifshitz-Gilbert (LLG) equation. Specifically, the evolution of the normalized magnetization $\textbf{m}$ in the YIG resonators is governed by the LLG equation,
\begin{equation}
    \frac{\partial \textbf{m}}{\partial t}=-\frac{\gamma}{1+\alpha^2}\textbf{m}\times\textbf{H}^\mathrm{eff}-\frac{\alpha\gamma}{1+\alpha^2}\textbf{m}\times (\textbf{m}\times \textbf{H}^\mathrm{eff}),
\label{Eq:LLG}
\end{equation}
\noindent where $\gamma$ is the gyromagnetic ratio and $\alpha$ is the effective magnetic damping coefficient. In this work, we consider that the total effective magnetic field is a sum of the following effective magnetic fields $\textbf{H}^\mathrm{eff} = \textbf{H}^\mathrm{anis}+\textbf{H}^\mathrm{bias}+\textbf{H}^\mathrm{dip}+\textbf{h}$. The magnetocrystalline anisotropy field $\textbf{H}^\mathrm{anis}$ is given by
\begin{equation}
H_i^\mathrm{anis}=-\frac{2}{\mu_0 M_s} \left[K_1(m_j^2+m_k^2)+K_2m_j^2m_k^2\right]m_i,
    \label{Eq:Hanis}
\end{equation}
\noindent where $\mu_0$ is the vacuum permeability, $M_s$ is the saturation magnetization, $K_1$ and $K_2$ are magnetocrystalline anisotropy coefficients, $i = x, y, z$, and $j \neq i$, $k \neq i, j$. For a thin magnetic slab (see main text) with uniform magnetization,  $\textbf{H}^\mathrm{dip} = -M_s(0, 0, m_z^0)$ describes the uniform magnetic dipolar coupling field inside the magnet, which is produced by $\textbf{m}_0=\textbf{m}(t=0)$ at the initial equilibrium state. Note that $\textbf{H}^\mathrm{dip}$ does not evolve with time since the magnetic dipole radiation is already considered via solving Maxwell’s equations, which will be discussed later. The \textbf{h} is the magnetic field component of the EM wave, and the EM wave in this work is produced by both the precessing $\textbf{m}(t)$ via magnetic dipole radiation and $\textbf{J}(t)$ via electric dipole radiation. The dynamics of the EM wave is governed by the Maxwell’s equations, and the two governing equations for the magnetic-field component $\textbf{h}$ and the electric-field component $\textbf{E}$ are given as
\begin{align}
\nabla\times \textbf{E} & =-\mu_0\left(\frac{\partial \textbf{h}}{\partial t}+M_s\frac{\partial \textbf{m}}{\partial t}\right),\\
\nabla\times \textbf{h} & =\varepsilon_0\varepsilon_r\frac{\partial \textbf{E}}{\partial t}+\textbf{J}(t),
\label{Eq:E}
\end{align}
\noindent where \textbf{m} is obtained by solving the LLG equation (Eq.\,\ref{Eq:LLG}), and $\varepsilon_0$ ($\varepsilon_r$) is the vacuum (relative) permittivity. Such two-way coupling between the EM wave and \textbf{m} is critical for accurately modeling the spatiotemporal evolution of systems containing interacting SSPP and magnon modes. \\

In our simulations, the material for the SSPP waveguide (the vertical parallel lines and the bottom horizontal line) is modeled as the perfect electric conductor (PEC) boundary condition, i.e., $E_i = E_j = 0$ on all surfaces that are in the $ij$ plane ($i = x, y, z$, and $j \neq i$), instead of real metals that have finite conductivities and nonzero absorptions. This will facilitate the convergence of the simulation without affecting the fundamental physical principles studied in this work. The PEC condition is also applied on both the top boundary of the whole simulation region. In addition, absorbing boundary condition (ABC) is applied on the two outer surfaces of the entire system that are normal to the $x$ axis [see Fig.\,\ref{fig_SM_config}(a)] to suppress the undesired reflections along the SSPP propagation direction, which can be expressed as $\frac{\partial E^\mathrm{EM}}{\partial x}=\frac{1}{v}\frac{\partial E^\mathrm{EM}}{\partial t}$ \cite{SM_Jin_Book_2010} with $v$ being the EM wave propagation speed in the material near the surface.\\

The whole simulation region is discretized into a system of cuboid cells $N_x\Delta x \times N_y \Delta y \times N_z \Delta z$ with cell size $\Delta x=\Delta y=10 \mu$m and $\Delta z=50 \mu$m, where $N_x = 1411$, $N_y=1$, and $N_z = 63$. Central finite difference is used for calculating spatial derivatives. All equations are solved simultaneously using the classical Runge-Kutta method for time-marching with a real-time step $\Delta t = 2\times 10^{-14}$ s. Specifically, the $\textbf{E}$ and $\textbf{h}$ components of the EM wave are discretized in conventional Yee grid \cite{SM_Taflove_book_2005} and Maxwell’s equations are numerically solved using finite difference time domain (FDTD) method. Also, each YIG resonator is described using one single computational cell under the macro spin model assumption, which is sufficient for describing the dominant Kittel mode magnon in the YIG. The material parameters of YIG used in the simulations are summarized below\cite{SM_Kamra_PRB_2015,SM_Azovtsev_PRB_2019,SM_Liu_NC_2018}: $\gamma = 27.86$ GHz/T, $\alpha = 8\times 10^{-5}$, $M_s = 0.14$ MA/m, and $K_1 = 602$ J/m$^{3}$. The relative permittivity of the entire simulation region is set as $\varepsilon_r=10$, which is close to that of the printed circuit board ($\varepsilon_r=9.8$) used in the experiments (see main text).

\begin{figure}[htb]
\includegraphics[width=0.7\linewidth]{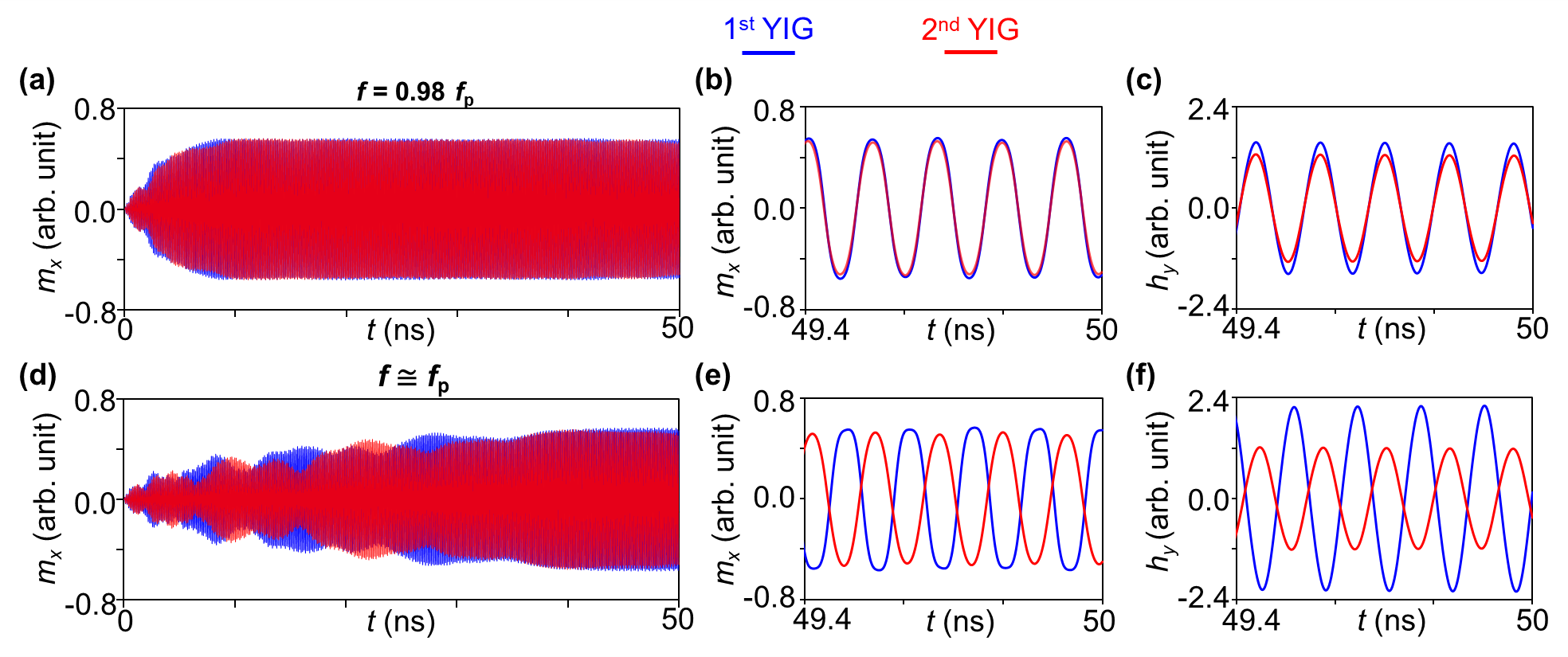}
\caption{(a) Evolutions of the Kittel mode magnons in the two YIG resonators shown in Fig.\,\ref{fig_SM_config}(a) within (a) $t =$ 0-50 ns and (b) the zoom-in view for $t =$ 49.4 – 50 ns under the excitation by continuous EM wave with a single frequency $f = 0.98 f_p$. (c) Evolution of the magnetic-field component hy of the EM wave at the positions of the two YIG resonators within $t =$ 49.4 – 50 ns with $f = 0.98 f_p$. (d)-(f), the corresponding data for the case of continuous wave EM excitation with a single frequency $f =0.9987f_p$. The $f_p = 7.8$ GHz is the cutoff frequency of the SSPP waveguide structure shown in Fig.\,\ref{fig_SM_config}(a). The bias magnetic field $H^\mathrm{bias}$ is tuned to ensure that the $f$ is equal to the FMR frequency of the magnons in both cases, where $H^\mathrm{bias}  = 4415.25$ Oe (4469.1 Oe) in the case of $f = 0.98f_p$ ($f = 0.9987f_p$).}
\label{fig_timetrace}
\end{figure}

To demonstrate the presence of the SSPP mode on the periodic metallic waveguide shown in Fig.\,\ref{fig_SM_config}(a), a broadband EM wave pulse is used as an input source for the SSPP waveguide. At this stage, only SSPP mode is considered, and therefore the $\textbf{H}^\mathrm{bias}$ is set to be 0 to disable the coupling between the magnon and the SSPP mode. Figure\,\ref{fig_SM_config} (b) presents the simulated temporal evolution of $h_y(t)$ of the SSPP mode inside the waveguide (taken at the position of the 2nd YIG resonator) after the injection of the EM wave. The EM wave is produced by applying a time-varying charge current pulse $\textbf{J}(t)$ on a $yz$-plane that is illustrated in Fig.\,\ref{fig_SM_config} (a), which is expressed as $J_z(t) = J_0 e^{(-t^2/2\sigma ^2)}\times \sin(2\pi f_0 t)$, where the pulse amplitude $J_0 = 10^9$ A/m$^{2}$, the center frequency $f_0 = 10$ GHz, and the Gaussian pulse duration $\sigma = 70$ ps. The simulated transmission spectrum of the SSPP waveguide is plotted in Fig.\,\ref{fig_SM_config} (c), which is obtained by performing Fourier transform on the $h_y(t)$ shown in Fig.\,\ref{fig_SM_config} (b). By comparing with the frequency spectrum of the excitation $J_z(t)$, one can find a cutoff frequency $f_p$ at 7.8 GHz, which matches the asymptotic frequency of the SSPP waveguide obtained from COMSOL Multiphysics using the same geometry.\\

Next, to demonstrate the frequency-controlled tuning of the relative phase the magnon modes in neighboring resonators, a continuous-wave (CW) charge current source $J_z(t) = J_0 \sin(2\pi ft)$ is injected into the system at the same position to directly drive the magnon mode. The source has an amplitude $J_0 = 10^9$ A/m$^{2}$ and its frequency is varied to match the magnon (FMR mode) frequency which is determined as $f_\mathrm{FMR}=\frac{\gamma}{2\pi}(H_z^\mathrm{bias} - M_s+\frac{2K_1}{\mu_0M_s})$ with initial $\textbf{m}^0 = (0, 0, 1)$ and $\textbf{H}^\mathrm{bias} = (0,0, H_z^\mathrm{bias})$. The detailed derivation of the $f_\mathrm{FMR}$ can be found in Ref.\,\cite{SM_Zhuang_NPJCM_2022}. Figure\,\ref{fig_timetrace} (a) shows the temporal evolution of the $m_x$ component of precessing magnetization in the two neighboring YIGs, which are driven by the CW EM field with a frequency $f$ slightly below the $f_p$ ($f=0.98 f_p$). It is evident that the magnon amplitudes first gradually increase and then reach the steady states. Figure\,\ref{fig_timetrace} (b) shows the zoomed-in time traces in the steady state ($t =$ 49.4 - 50 ns), which follow the evolution of magnetic-field component $h_y(t)$ of the EM waves at the same position [Fig.\,\ref{fig_timetrace} (c)] with a constant phase delay. Clearly at this frequency the two magnon modes have nearly identical phases. As the driving frequency $f$ (and the magnon frequency) increases and approaches the cutoff frequency ($f=0.9987f_p$), it takes longer time for the magnons to reach the steady state [Fig.\,\ref{fig_timetrace} (d)], which can be explained by the reduced group velocity of the SSPP mode. More importantly, the relative phase of the magnon modes in the two neighboring YIG resonators is approaching the theoretical limit of $\pi$, as shown in Fig.\,\ref{fig_timetrace} (e). This is a result of the nearly $\pi$ phase difference between the $h_y(t)$ fields that drive the magnons in the two magnonic resonators near $f_p$ [Fig.\,\ref{fig_timetrace} (f)], which corresponds to the boundary of the first Brillouin zone of the SSPP dispersion where $\beta=\pi/d$.

\section{Device Preparation}
The YIG chip is fabricated on a 200-nm YIG film epitaxially grown on a 500-$\mu$-thick GGG substrate using photolithography and dry etching method. The SSPP waveguides, CPWs, and microstrips are fabricated on printed circuit boards with a substrate dielectric constant of 9.8. On all these circuits, the backside ground planes are all kept to be consistent. The YIG chip is bonded to the circuits using GE varnish, after the YIG chip is flipped over the circuit (with the YIG side facing the circuit). Since the YIG/GGG chip is transparent, we used a microscope to carefully adjust its position on the circuit.\\

A permanent magnet is used to provide the bias magnetic field for the magnon excitation. In principle, the bias magnetic field can be along any directions to have magnon-photon (SSPP) coupling, because the microwave magnetic field associated with the propagating SSPPs are mainly along both $y$ and $z$ directions, i.e., transverse directions with respect the propagation direction of the SSPP modes, and their amplitudes are similar to each other. Therefore, when the bias field is applied along $y$ ($z$) direction, it will be orthogonal with the $z$ ($y$) component of the microwave magnetic field. If the bias field is applied along $x$ direction, it will be orthogonal with both $y$ and $z$ components. In our experiment, the bias field is applied along $z$ direction, i.e., the out-of-plane direction, because of it is more convenient to implement for the permanent magnet and the planar device geometry used in our case.\\

The SSPP waveguide, CPW, and microstrips are all fabricated from the same type of printed circuit boards. The substrate has a thickness of 0.5 mm with a dielectric constant of 9.8, and the copper thickness is 35 um. All the three types of waveguides have the ground plane on the backside for consistency. The width of the signal lines for the microstrip is 3 mm, which is the same as the total width of the SSPP waveguide. The center line of the CPW has a width of 100 $\mu$m, with a gap of 500 $\mu$m from the ground.

\section{Transmission spectrum}

\begin{figure}[hbt]
\includegraphics[width=0.75\linewidth]{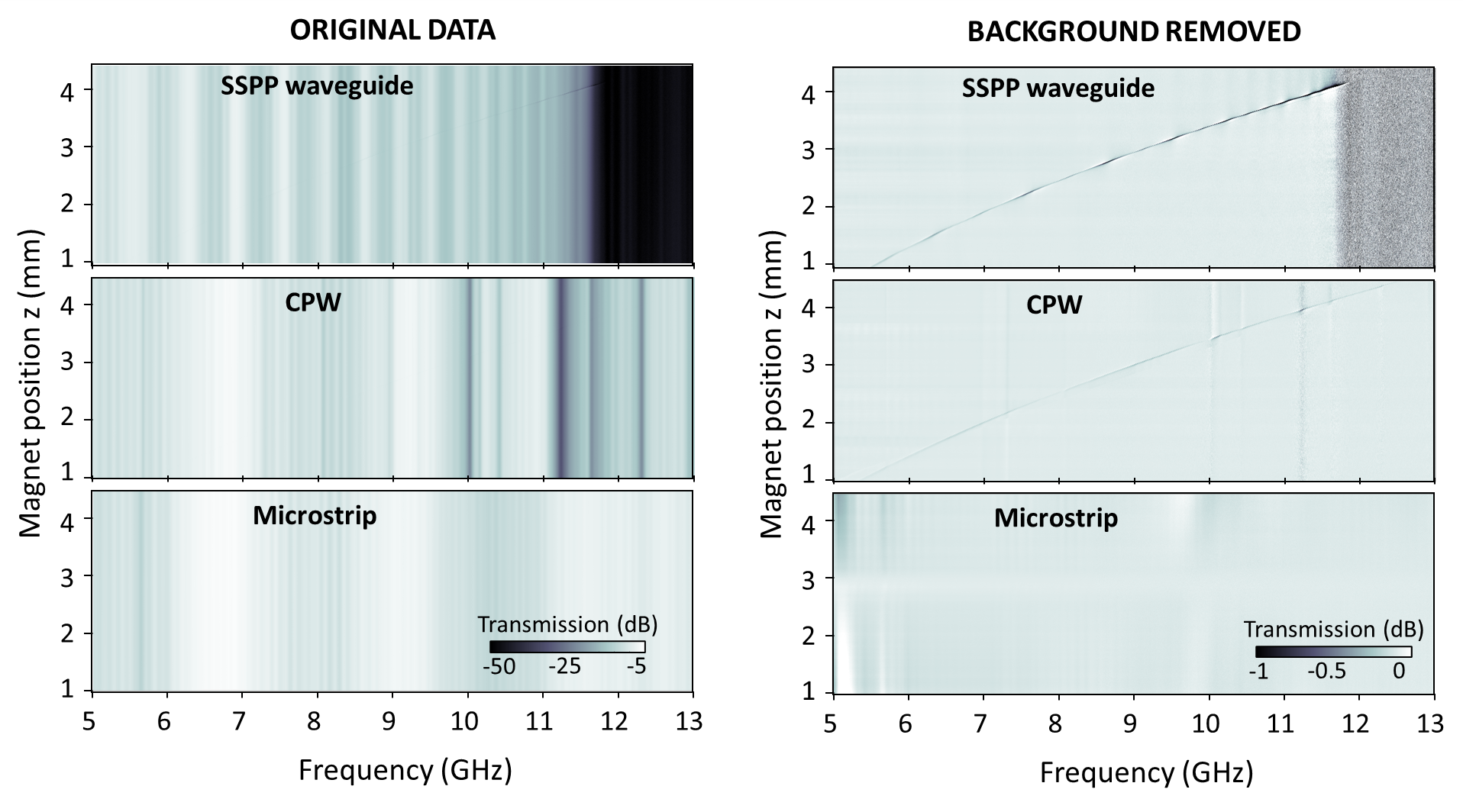}
\caption{Original data vs background-removed data as shown in Fig.2 of the main text.}
\label{trans}
\end{figure}

The devices are characterized by transmission measurements using a vector network analyzer. The enhanced coupling between the SSPP waveguide and magnonic resonator (as compared with the case of CPW or microstrip) is more clearly shown by the original data of the measured transmission [Fig.\,\ref{trans}] that correspond to the background-removed data in Fig.\,2(c) of the main text. The plotted transmission range is from -50 dB to -5 dB to accommodate the abrupt transmission change on the SSPP waveguide around the cutoff. With such a large dynamic range, the magnon mode is still visible near the cutoff frequency, thanks to the enhanced SSPP-magnon coupling. On the contrary, the weak magnon signals in the CPW transmission become completely invisible because of their small ($<1$ dB) extinction, which can only be seen after the background removal. Note that in the background-removed spectra, even though the magnon modes show similar colors in the SSPP waveguide and CPW spectra, their extiction ratios are quite different. This is because within the small colorbar range (-1 dB to 0 dB) which is necessary to show the weak magnon modes in the CPW spectra, the magnon resonances in the background-removed spectra of the SSPP waveguide are already saturated, limiting their color contrast in the plots.\\

\begin{figure}[hbt]
\includegraphics[width=0.75\linewidth]{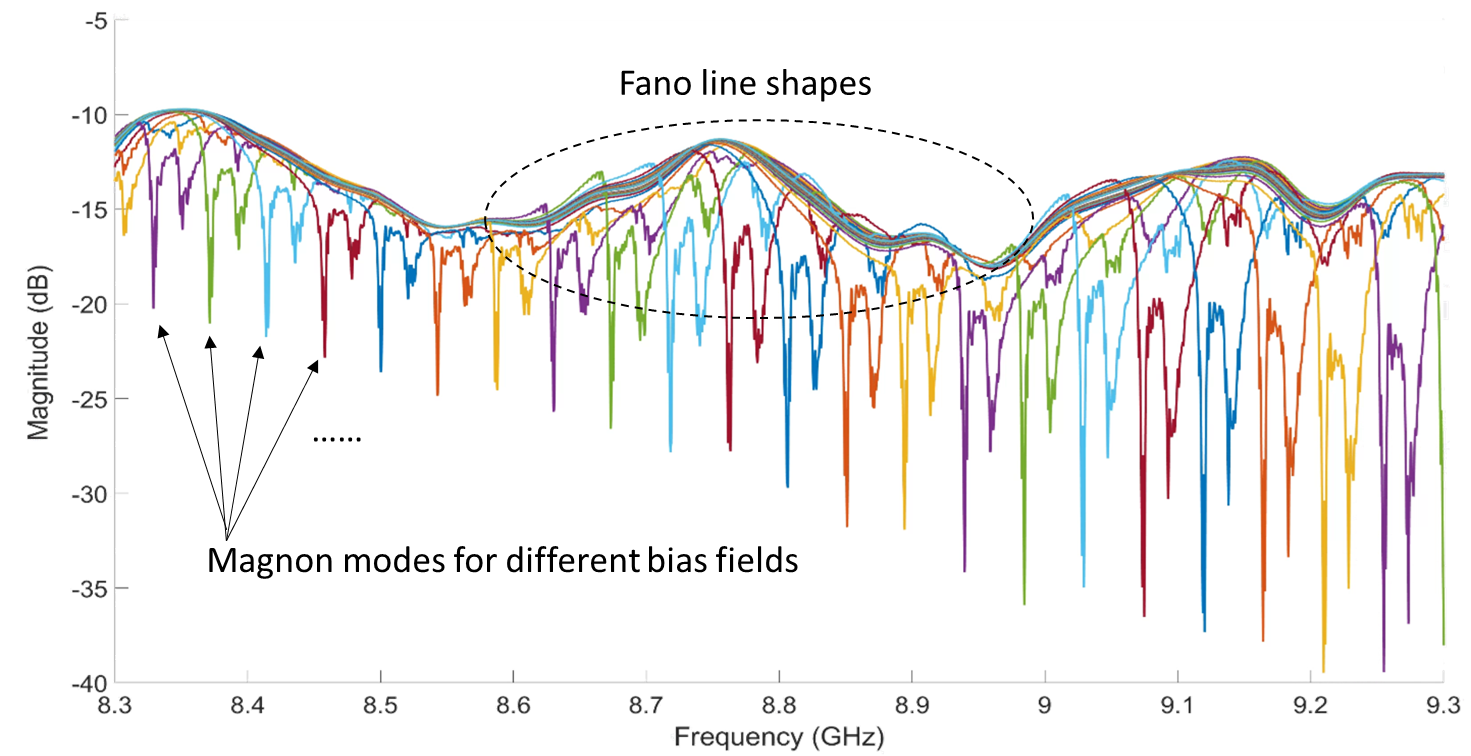}
\caption{Zoom-in of the measured device transmission spectra around 8.5-9 GHz for different bias magnetic fields.}
\label{spectra}
\end{figure}

The unprocessed transmission spectra are plotted in Fig.\,\ref{spectra} for different bias magnetic fields. It can be seen that when the magnon mode is tuned to the 8.5-9 GHz range, their resonances interfere with the slow oscillation of the background signal and form Fano line shapes, which degrade the accuracy of the numerical fitting for the magnon linewidth and the extraction of their extinction ratios. Consequently, this leads to the fluctuation of the coupling strengths calculated using these quantities between 8.5-9 GHz in Fig.\,2(c) of the main text.

\bibliography{Supplemental}
